\documentclass[10pt]{article}
\usepackage{multicol,epsfig,citesort}
\begin{document}

\title{Metastability and nucleation in the dilute fluid phase of
a simple model of globular proteins}

\author{{\bf Richard P. Sear}\\
~\\
Department of Physics, University of Surrey\\
Guildford, Surrey GU2 5XH, United Kingdom\\
email: r.sear@surrey.ac.uk}

\date{}

\maketitle

\begin{abstract}
The dilute fluid phase of model globular proteins is studied.
The model possesses a fluid-fluid transition buried within
the fluid-crystal coexistence region, as do some globular proteins.
If this fluid-fluid transition is not buried deep inside the
fluid-crystal coexistence region the crystalline phase
does not nucleate within the dilute fluid.
We link this lack of nucleation of the crystal to the interactions
in our model and speculate that similar interactions between
globular proteins are responsible for the difficulty found
in crystallising many globular proteins.
\end{abstract}



\section{Introduction}

Metastability is the persistence for a long time of a phase which
is not the equilibrium phase. It can be both a blessing and a curse.
In protein solutions it is a curse. Protein crystals are required
for X-ray crystallography to determine their full structure
\cite{durbin96,chayen99} but protein solutions at concentrations well in
excess of the solubility of the crystalline phase are often
stable essentially indefinitely; the rate of nucleation of the
crystalline phase is essentially zero.
Here, we consider a crude model
of a globular protein and we find that depending on the
parameters of the model, the dilute fluid phase may be stable
indefinitely with respect to crystallisation. If the solution is
cooled at some low density, it is stable with respect to crystallisation
down to temperatures at which the solution undergoes a fluid-fluid
transition. This transition has been observed in protein
solutions \cite{broide91,muschol97}.
The agreement in the phase and nucleation behaviour between the
simple model and experiment is encouraging. It is clear what
underlies the behaviour of the model and we may hope that similar
physics underlies the behaviour of globular proteins.
The fluid is metastable for a wide range of parameters and
temperatures because as the attractions are directional
and short-ranged the crystal is only stable when these attractions
are strong relative to the thermal energy $kT$. These strong attractions
mean that the interfacial tension between the dilute fluid and crystalline
phases is high and it is this that inhibits crystallisation.
This suggests that to increase the nucleation rate the attractions
should be modified to become more like that in argon or other
simple atoms and molecules, i.e., to become less anisotropic and
longer ranged. The phase diagram will correspondingly become
more like that of simple atomic fluids.

The interactions between globular
proteins are rather poorly understood but it seems clear
that many of the attractive interactions are
directional and quite short-ranged \cite{durbin96,visser92,neal99,haas99};
two protein molecules must not only be close
to each other to attract each other but they must also be correctly oriented.
An example is the attraction between hydrophobic patches on
the surfaces of globular proteins; only
if the proteins are oriented so that these parts of their
surfaces face each other is there an attraction.
So, our model, specified in section 2, contains directional
attractions; in fact for simplicity it contains only directional
attractions.
The model was introduced by us and its bulk phase behaviour
calculated in Ref. \cite{searprot}. Here, we
carefully define metastability and derive an approximate theory
to tell us when the dilute fluid is metastable and when nucleation
occurs. We then present and discuss results, and finish
with a conclusion.

\section{Model}

Our model is exactly the same as in Ref. \cite{searprot}.
The potential is a pair potential $\phi$ which is a sum of two
parts: a hard-sphere repulsion, $\phi_{hs}$, and a set of sites
which mediate short-range, directional attractions. There
are $n_s$ sites, where $n_s$ is an even integer.
In order to keep the model as simple as possible there are no
isotropic attractions and all the directional
attractions are of the same strength.
The sites come in pairs: a site on one particle binds only to the
other site of the pair on another particle.
The two sites of a pair are numbered consecutively
so that an odd-numbered site, $i$, binds only to the even-numbered site,
$i+1$. This is the only interaction between the sites, an odd-numbered
site, $i$, does not interact at all with sites other than the $(i+1)$th
site. The orientation of site number $i$
is specified by means of a unit vector ${\bf u}_i$.
We can write the interaction potential between a pair of particles as
\begin{equation}
\phi(r_{12},\Omega_1,\Omega_2)=\phi_{hs}(r_{12})
+\sum_i^{'}
\left[  \phi_{ii+1}(r_{12},\Omega_1,\Omega_2)
+  \phi_{ii+1}(r_{12},\Omega_2,\Omega_1)\right],
\label{pot}
\end{equation}
where the dash on the first sum denotes that it is restricted to
odd values of $i$. The interactions between the sites
on the two particles are $\phi_{ii+1}(r_{12},\Omega_1,\Omega_2)$, which
is the interaction between site $i$ on particle 1 and site $(i+1)$ on
particle 2, and $\phi_{ii+1}(r_{12},\Omega_2,\Omega_1)$, which
is the interaction between site $i$ on particle 2 and site $(i+1)$ on
particle 1. These are functions of $r_{12}$, $\Omega_1$ and $\Omega_2$,
which are the scalar
distance between the centres of particles 1 and 2, the orientation
of particle 1 and the orientation of particle 2, respectively.
The particle is rigid, but not axially symmetric, so
its position is completely specified by the position of its centre and
its orientation $\Omega$, which may be expressed in terms of the
three Euler angles.

The hard-sphere potential, $\phi_{hs}$, is given by
\begin{equation}
\phi_{hs}(r)=
\left\{
\begin{array}{ll}
\infty & ~~~~~~ r \le \sigma\\
0 & ~~~~~~ r > \sigma\\
\end{array}\right. ,
\label{hspot}
\end{equation}
where $\sigma$ is the hard-sphere diameter. The conical-site
interaction potential $\phi_{ii+1}$ is given by \cite{jackson88}
\begin{equation}
\phi_{ii+1}(r_{12},\Omega_1,\Omega_2)=
\left\{
\begin{array}{ll}
-\epsilon & ~~~~~~
r_{12}\le r_c ~~~~~ \mbox{and} ~~~~~ \theta_{1i}\le\theta_c ~~~~~ \mbox{and}
 ~~~~~ \theta_{2i+1}\le\theta_c\\
0 & ~~~~~~ \mbox{otherwise}\\
\end{array}\right. ,
\label{sitepot}
\end{equation}
where $\theta_{1i}$ is the angle
between a line joining the centres of the two particles and the
unit vector ${\bf u}_i$ of particle 1, and
$\theta_{2i+1}$ is the angle
between a line joining the centres of the two particles and the
unit vector ${\bf u}_{i+1}$ of particle 2.
The conical-site potential depends on two parameters:
the range, $r_c$, and the maximum angle at which a bond is
formed, $\theta_c$. Of course, as the attractions are directional,
$\theta_c$ will be small, no more than about $30^{\circ}$.
The attractions are also short ranged, $r_c$ no more than 10\%
larger than $\sigma$.

The angles between the site orientations, the vectors ${\bf u}_i$
will determine which crystal lattice is formed. 
For simplicity, we will take the
sites to be arranged such that they are compatible with a simple cubic
lattice. Then if we express the unit vectors ${\bf u}_i$ in
Cartesian coordinates, $(x,y,z)$, then when we have four sites, $n_s=4$,
the set of vectors
${\bf u}_1=(1,0,0)$, ${\bf u}_2=(-1,0,0)$, ${\bf u}_3=(0,1,0)$ and
${\bf u}_4=(0,-1,0)$ would describe our model.
For six sites then we add two additional sites at orientations
${\bf u}_5=(0,0,1)$ and ${\bf u}_6=(0,0,-1)$.

Later on when we discuss the metastable fluid we will discuss rates.
In order to do this we let $\tau$ be the characteristic time for the
dynamics in the dilute fluid. We will not need to specify $\tau$
exactly but it is of order the time a molecule takes to diffuse
the average separation between the molecules.

\section{Theory for the bulk phases}

The free energies of the fluid and bulk phases were derived
in our previous paper, Ref. \cite{searprot}, so we only outline
their derivations here.  See Ref. \cite{searprot} for details.

\subsection{Theory for the fluid phase}

The theory for the fluid phase of particles interacting via a hard-core
and directional attractions mediated by sites is well
established \cite{wertheim,jackson88,chapman88}.
Our theory is based on the generalisation of
Chapman, Jackson and Gubbins \cite{jackson88,chapman88}
of Wertheim's perturbation theory \cite{wertheim}.
The perturbation theory gives for the Helmholtz free energy per
particle of the fluid phase, $a_f$, \cite{jackson88}
\begin{equation}
\beta a_f (\eta,T)=\beta a_{hs}(\eta) + n_s\beta\Delta a(\eta,T),
\label{af}
\end{equation}
where $a_{hs}$ is the Helmholtz free energy per particle of a fluid of
hard spheres, and
$\Delta a$ is the change in free energy per bonding site
due to bonding,
\begin{equation}
\beta\Delta a=\ln X+\frac{1}{2}(1-X).
\label{delta}
\end{equation}
We use an accurate expression
derived from the equation for the pressure of Carnahan and
Starling \cite{carnahan69,hansen86} for $a_{hs}$. The volume
fraction $\eta=(N/V)(\pi/6)\sigma^3$ is a reduced density, it
is the fraction of the solution's volume occupied by the molecules.
$N$ and $V$ are the number of molecules and the volume, respectively.
$\beta=1/kT$, where $k$ is Boltzmann's constant and $T$ is the
temperature. $X$ is the fraction of sites which are {\em not}
bonded to another site. As all site-site interactions are equivalent
the fraction of each type of site which is not bonded is the same.
The fraction of sites which are bonded and the fraction
which are not bonded must, of course, add up to one. Thus we can
simply write down a mass-action equation for $X$, \cite{jackson88}
\begin{equation}
1=X+\rho X^2 Kg_{hs}^c(\eta)\exp(\beta\epsilon),
\label{ma}
\end{equation}
where 
$g_{hs}^c$ is the contact value of pair distribution function
of a fluid of hard spheres, and $\rho=(N/V)\sigma^3$.
The volume of phase space (both translational and orientational
coordinates) over which a bond exists is $K$ \cite{jackson88},
\begin{equation}
K=\pi\sigma^2(r_c-\sigma)(1-\cos\theta_c)^2.
\label{kdef}
\end{equation}
The mass-action equation, Eq. (\ref{ma}), is a quadratic equation
for $X$ and it can be solved for $X$. Inserting this solution in Eq.
(\ref{delta}) and then the result into Eq. (\ref{af})
yields the Helmholtz free energy as a function of density and temperature.
The state of our single component fluid is specified by the ratio of the
site energy to the thermal energy, $\beta\epsilon$,
and the volume fraction, $\eta$.
Note that Eq. (\ref{ma}) is not quite the same as the equivalent
equations in Refs. \cite{wertheim,jackson88,chapman88}.
In those references $\exp(\beta\epsilon)$ is replaced by
$\exp(\beta\epsilon)-1$. As $\beta\epsilon$ is quite large,
five or more, the difference between the two is very small.
Also, our $K$ is $4\pi$ times the $K_{AB}$ of Ref. \cite{jackson88}.

The second virial coefficient $B_2$ was obtained in Ref. \cite{searprot}.
It is
\begin{equation}
B_2= B_2^{hs} - \frac{n_s}{2}K\exp(\beta\epsilon),
\label{b2}
\end{equation}
where $B_2^{hs}=(2\pi/3)\sigma^3$ is
the second virial coefficient of hard spheres.

\subsection{Theory for the crystalline phase}

At low temperature, crystallisation is driven by the attractive interactions,
not packing effects as it is with hard spheres.
In Ref. \cite{searprot} we used
a cell theory to describe the
free energy of the crystalline phase of our model \cite{buehler51,sear98}. 
The theory is a low-temperature theory and we will use
it only at low temperatures.
Vega and Monson \cite{vega98} used a cell theory to describe
the solid phase of a very similar model, a simple model of
water. They avoid a couple of the approximations used here at the
cost of not having an analytical free energy.
Within a cell theory for a solid phase,
the Helmholtz free energy per particle, $a_s$, is given by
\begin{equation}
\beta a_s(\eta,T)=-\ln q_P
\label{adef}
\end{equation}
where $q_P$ is the partition function of a single particle
trapped in a cage formed by the requirements that all its $n_s$
sites bond to neighbouring particles, and that its hard core
not overlap with any of these neighbours. If the lattice
constant is $b$, then the particle can move a distance
distance $b-\sigma$ in the direction of any of its neighbours,
without overlapping with the neighbour.
In order for the bonds to not be broken the particle
must always be within $r_c$ of the surrounding particles.
This fixes the lattice constant, $b$, at a little less than $r_c$.
It is a little less as when the particle moves off the lattice
site it will be moving towards some of its neighbours and
away from others. Thus it can explore regions where it is further
than $b$ from some of its neighbours.
The exact value of the maximum lattice constant for which
the particle can move about, constrained by the hard-sphere
interactions, without breaking any bond, is difficult to
estimate; as is the volume available to the centre of mass of
the particle \cite{buehler51}.
Therefore, we approximate the lattice constant $b$ by
$r_c$ and the volume to which the particle is restricted by
$(r_c-\sigma)^3$.
The requirement that no bonds be broken also severely restricts
the orientations of the particle. When a non-axially symmetric
particle is free to rotate it explores an angular phase space of
$8\pi^2$. However, in the crystal its rotations will be restricted
to those which are small enough not to violate the requirement
that the orientations of its site vectors are within $\theta_c$
of the lines joining the centre of the particle with the those
of the neighbouring particles. Again the exact value
of angular space available to the particle is complex, and
it also depends on the position of the particle. We
approximate this angular space by assuming that each of the
three angular degrees of freedom can vary independently
over a range of $2\theta_c$. The normalised angular space
available to a particle in the solid phase is then
$(2\theta_c)^3/8\pi^2=\theta_c^3/\pi^2$.
The energy per particle is, of course, $-(n_s/2)\epsilon$, and so
the partition function, $q_P$, is then just the volume available to the
centre of mass of the particle times
the angular space available times $\Lambda^{-1}\exp[(n_s/2)\beta\epsilon]$,
where $\Lambda^{-1}$ is the integral over the momentum degrees of freedom.
Thus, we have for $q_P$,
\begin{equation}
q_P=v_P\Lambda^{-1}\exp\left(\frac{n_s}{2}\beta\epsilon\right),
\label{qp}
\end{equation}
where
\begin{equation}
v_P=(r_c-\sigma)^3\left(\frac{\theta_c^3}{\pi^2}\right).
\label{vp}
\end{equation}
Inserting Eq. (\ref{qp}) for $q_P$ into Eq. (\ref{adef}),
\begin{equation}
\beta a_s=-\ln\left(v_P/\Lambda\right)
-\frac{n_s}{2}\beta\epsilon
=\beta \mu_s.
\label{mus}
\end{equation}
This is the free energy at a lattice constant of $r_c$.
The maximum possible density of a simple-cubic lattice is
when the lattice constant $b=\sigma$, then the density is
$\sigma^{-3}$.
This density corresponds to a volume fraction $\eta=\pi/6$.
When the lattice constant is $r_c$, the density is  
$r_c^{-3}$ and the volume fraction is
$(\pi/6)(\sigma/r_c)^3$.

We are interested in finding coexistence between the crystal phase and
the fluid phase at low temperature, when our assumption that
no bonds are broken in the solid phase will be accurate. Then the
pressure at coexistence will be low and the solid will be near
its minimum possible density, $r_c^{-3}$.
The chemical potential $\mu_s=a_s+p_s/\rho$ where $p_s$ is the
pressure and $\rho$ is the density. At low pressure $p_s/\rho$
contributes a negligible amount to the chemical potential, which
enables us to equate $a_s$ and $\mu_s$ as we have done in Eq. (\ref{mus}).
The coexisting fluid density at the fluid-solid transition
is then found by equating the chemical potentials in the two phases.
The density of the coexisting solid phase, when the temperature is low enough
that solidification is driven by the attractive interactions
not packing effects, is assumed constant at $r_c^{-3}$.
See Ref. \cite{searprot} for details.

\section{Crystalline clusters}

We derive a simple but rather crude approximation for the
equilibrium density
of crystalline clusters in a dilute fluid. The
approximations used are similar in spirit
to our calculation of the interfacial tension between the crystal
and dilute fluid phases of the spheres with a short-range isotropic
attraction \cite{sear99,sear99b}.
We will assume that the interface
between the cluster
and the surrounding dilute fluid is sharp and that the interaction
between the crystalline cluster and the surrounding fluid is weak.
Both these assumptions are reasonable if the fluid is dilute but not if
it is dense or near a fluid-fluid critical point \cite{tenwolde97}.
Thus we will only be able to predict the densities of crystalline
clusters and therefore the nucleation rate of the crystalline
phase in the dilute fluid.

We require the density of crystalline
clusters of $n$ particles, $\rho_c(n)$, in a dilute gas.
To find this we start from the $n$-particle distribution function,
$\rho^{(n)}(1\ldots n)$
in the grand-canonical ensemble \cite{hansen86,stell64}
\begin{equation}
\rho^{(n)}(1\ldots n)=\frac{\sum_N\frac{z^N}{(N-n)!}
\int {\rm d}(n+1)\ldots {\rm d }(N)\exp\left(-\beta U\right)}
{\sum_N\frac{z^N}{N!}
\int {\rm d}(1)\ldots {\rm d}(N)\exp\left(-\beta U\right)},
\label{rhon}
\end{equation}
where $(i)$ is a compact form for the positional, ${\bf r}_i$,
and orientational, $\Omega_i$, coordinates of molecule $i$,
$(1\ldots n)$ indicates that $\rho^{(n)}$ is a function
of the set of $n$ coordinates of the $n$ molecules. $U$ is the
total energy of the fluid and depends on all $N$ coordinates.
\begin{equation}
z=\Lambda^{-1}\exp(\beta\mu)
\label{activity}
\end{equation}
is the activity.

Equation (\ref{rhon}) gives the density of an $n$-tuple of particles with
coordinates $(1\ldots n)$ in the fluid. We want the density
of an $n$-tuple of molecules which are in a configuration which
is compatible with the $n$ molecules being part of a single compact
crystalline cluster.
Therefore we integrate over all the positions
of the $n$ particles which are consistent with the $n$
particles forming a crystalline cluster, and over no other positions.
Integration over all $n$ coordinates
will give us the total number of crystalline clusters, to obtain the number
density $\rho_c(n)$ (here $(n)$ indicates the dependence of
$\rho_c$ on $n$ the number of molecules in the cluster,
not that $\rho_c$ depends on the coordinates of the $n$th molecule)
we divide by the volume,
\begin{equation}
\rho_c(n)=\frac{1}{n!V}
\int' {\rm d}(1)\ldots {\rm d }(n)\rho^{(n)}(1\ldots n),
\label{rhocdef}
\end{equation}
where the dash on the integration sign indicates that the integration
is restricted to those configurations of the $n$ particles which
are consistent with them forming a cluster. The factor of $1/n!$
is present
because the particles are indistinguishable and so the integral
integrates over configurations which differ only by the
exchange of indistinguishable particles.

As we are assuming that the cluster is in an ideal gas
Eq. (\ref{rhon}) simplifies as we set the energy of interaction
to be zero except for the energy of interaction between the
$n$ particles in the cluster. Then the integral in the
denominator of Eq. (\ref{rhon}) is simply $V^N$ and that in the
numerator is $V^{N-n}\exp(-\beta u(1\ldots n))$, where
$u(1\ldots n)$ is the energy of interaction of $n$ molecules.
So Eq. (\ref{rhon}) becomes
\begin{equation}
\rho^{(n)}(1\ldots n)=
\frac{\sum_N\frac{z^NV^{N-n}}{(N-n)!}
\exp\left(-\beta u(1\ldots n)\right)}
{\sum_N\frac{z^NV^N}{N!}}.
\end{equation}
Substituting this in Eq. (\ref{rhocdef}),
\begin{equation}
\rho_c(n)=
\frac{\sum_N\frac{z^NV^{N-n}}{(N-n)!}
\int'{\rm d}(1)\ldots {\rm d }(n)\exp\left(-\beta u(1\ldots n)\right)}
{n!V\sum_N\frac{z^NV^N}{N!}}.
\end{equation}
We can take $z^n$ times the integral out of the sum in the numerator
leaving the sum in the numerator identical to that in the denominator.
They cancel leaving
\begin{equation}
\rho_c(n)=
\frac{z^n}{Vn!}\int'{\rm d}(1)\ldots {\rm d }(n)
\exp\left(-\beta u(1\ldots n)\right).
\label{rhocexp}
\end{equation}
The density of crystalline clusters of $n$ molecules in an ideal
gas is simply $z^n/Vn!$ times the configurational integral
of $n$ molecules in a cluster.

As in the cell theory for a bulk crystal we factorise the integration of
Eq. (\ref{rhocexp}) into a product of $n$ integrals and delete
the factor of $1/n!$ as once the molecules are restricted to lie in cells they
are distinguishable. Now, one of the $n$ integrations is over the whole
volume $V$ of the fluid, the other $(n-1)$ are just over the rattling motion
as in the bulk and they each give a factor of $v_P$.
The energy is taken to be the
ground state energy as in the bulk and so is $-nn_s\epsilon/2+u_s(n)$
where $u_s$ is the increase in energy due to broken bonds at the surface of the cluster. So, we have that Eq. (\ref{rhocexp}) becomes
\begin{equation}
\rho_c(n)=
z^nv_P^{n-1}\exp\left[\frac{nn_s}{2}\beta\epsilon-\beta u_s\right].
\label{rhoc2}
\end{equation}
The spheres at the faces of the cluster do not interact with the full
$n_s$ other spheres and this increases the energy of a cluster.
If we assume that the cluster of $n$ molecules is cubic then it has 6 faces,
each of area $n^{2/3}\sigma^2$, i.e., with $n^{2/3}$ molecules in each face.
For $n_s=6$ there are sites pointing in all 6 directions and
a sphere at any of the 6 faces but not at an edge or corner
has one bond broken. So assuming that the cluster is cubic,
neglecting the fact some spheres are at edges and some at
corners and therefore have 2 or 3 bonds
not 1 bond broken and treating $n$ as a continuous variable,
results in the approximation that there are $6n^{2/3}$ bonds broken on
the surface of the cluster. Each broken bond costs an energy
$\epsilon/2$ --- the energy of a bond
is $\epsilon$ with $\epsilon/2$ assigned to each of the
two particles forming the bond. Thus, for $n_s=6$,
the increase in energy $u_s=3n^{2/3}\epsilon$.
With this expression for $u_s$ Eq. (\ref{rhoc2}) becomes
\begin{equation}
\rho_c(n)=
z^nv_P^{n-1}\exp
\left[\frac{n_sn}{2}\beta\epsilon-3n^{2/3}\beta\epsilon\right]
~~~~~~~~~~n_s=6.
\end{equation}
The approximation $u_s=3n^{2/3}\epsilon$
becomes worse as $n$ decreases but it is
never seriously wrong. Indeed for the smallest cluster we consider,
that of 8 spheres,
there is cancellation of errors and there are exactly 
$6\times8^{2/3}=24$ bonds broken. For $n=9$ we predict
26.0 bonds broken when in fact there 28 broken bonds but this
is not a large error.
For $n_s=4$ only 4 of the 6 faces involve broken bonds, because there are no
attraction sites on 2 faces. So, instead the energy cost is only
two thirds that for 6 sites and the increase in energy is
$2n^{2/3}\epsilon$. For $n_s=4$ or 6 the increase in energy is given
by $(n_s/2)n^{2/3}\epsilon$.

So far we have assumed that the cluster does not interact with any of the
surrounding spheres. This is reasonable for a very dilute fluid. However
for a fluid which is not very dilute and is at low temperature,
spheres in the surrounding fluid will tend to bond to the spheres
in the faces of the cluster. We can take this into account approximately
by treating the sites on the faces of a crystalline cluster as if
they were sites in the fluid, then for each site there is a free
energy change given by Eq. (\ref{delta}) --- which reflects the fact
that it can bond to one of the surrounding spheres. The
change to the configurational integral is then of course
$\exp(-\beta\Delta a)$ per surface site.
Then the configurational integral is
\begin{equation}
\rho_c(n)=z^nv_P^{n-1}
\exp\left[\frac{n_sn}{2}\beta\epsilon-
n_sn^{2/3}\left(\frac{\beta\epsilon}{2}+
\beta\Delta a\right)\right],
\label{rhocfin}
\end{equation}
where $X$ in Eq. (\ref{delta}) for $\Delta a$
is the same as in the surrounding fluid.

\section{Metastability and nucleation}

Consider the density of clusters $\rho_c(n)$ of Eq. (\ref{rhocfin}).
For large $n$, $\rho_c$ is dominated by the part
$(zv_P\exp[(n_s/2)\beta\epsilon])^n$ as the other
parts vary only as the $n^{2/3}$ power or are constants.
Using Eqs. (\ref{mus}) and (\ref{activity}), we obtain
\begin{equation}
zv_P\exp[(n_s/2)\beta\epsilon]=\exp(\beta\mu)\exp(-\beta\mu_s).
\label{bigger}
\end{equation}
But as we are within the fluid-crystal coexistence region the chemical
potential of the crystal $\mu_s$,
is less than that of the fluid phase, $\mu$. So, Eq. (\ref{bigger})
is greater than 1 and hence $\rho_c(n)$ diverges as $n\rightarrow\infty$.
This is actually an automatic consequence of the fact that the crystal
is more stable than the fluid.

So, our Eq. (\ref{rhocfin}) predicts that in the
fluid there are high densities of large crystalline clusters.
This is of course not what is observed in a metastable fluid.
This is because our calculation of Eq. (\ref{rhocfin}) assumed
that the densities of all clusters were at equilibrium,
whereas in a metastable fluid the system is by definition not
at equilibrium.
In order to describe a metastable fluid, a fluid which is out
of true equilibrium, we must apply a
constraint; see
Refs \cite{debenedetti,penrose71,stillinger88,corti97,corti98}
for definitions and discussions of the application of
constraints to study metastable fluids.
This constraint
must eliminate the large crystalline clusters to leave us with a
fluid. We choose the constraint which eliminates
all clusters above a size $n_{min}$:
\begin{equation}
\rho_c(n)=0~~~~~~~~~~~~~~~~~n>n_{min},
\label{con}
\end{equation}
where $n_{min}$ is defined by
\begin{equation}
\rho_c(n_{min})= \min_n\left\{\rho_n\right\},
\end{equation}
i.e., $n_{min}$ is the number of molecules in the cluster with
the lowest density, as predicted by Eq. (\ref{rhocfin}).
So, our constrained distribution of cluster densities is
\begin{equation}
\rho_c(n)= \left\{
\begin{array}{ll}z^nv_P^{n-1}
\exp\left[\frac{n_sn}{2}\beta\epsilon-
n_sn^{2/3}\left(\frac{\beta\epsilon}{2}+
\beta\Delta a\right)\right],
 & ~~~~~~ n\le n_{min} \\
0 & ~~~~~~ n > n_{min} \\
\end{array}\right. .
\label{rhocon}
\end{equation}
We set the constraint so as to eliminate all clusters above the size
$n_{min}$ because this constraint is in a specific sense the least
restrictive. It is the least restrictive because if we start with the
constrained equilibrium distribution of clusters, which is given by
Eq. (\ref{rhocon}) and then remove the constraint, i.e., allow
clusters with $n>n_{min}$ to form, then the initial rate at which
these clusters with $n>n_{min}$ form is minimised.
This assumes that clusters only grow one molecule at a time; that
a cluster with $(n+1)$ molecules is formed by a cluster of
$n$ molecules adsorbing an additional molecule.
This is a reasonable
assumption in a dilute fluid in which the density of single
molecules is much larger than the density of clusters of 2 or more
molecules.
With this assumption of growth one molecule at a time the initial
rate at which clusters with $n>n_{min}$ appear is just equal
to the rate at which clusters of $n_{min}$ molecules acquire an additional
molecule to become clusters of $(n_{min}+1)$ molecules, which is
approximately
\begin{equation}
\mbox{rate}\sim\rho_c(n_{min})\tau^{-1}.
\label{rate}
\end{equation}
Therefore, with our choice of constraint the initial rate at which the
distribution of clusters changes when the constraint is removed is
minimised. This is what we meant by the constraint being least
restrictive.
When the constraint is removed the distribution will tend towards
the equilibrium one with its crystalline-cluster densities which diverge
in the $n\rightarrow\infty$ limit, i.e., the fluid will crystallise.
If we neglect the fact that not all the clusters with $n_{min}$
molecules which gain an extra molecule will grow all the way into
a crystallite, then the rate of nucleation of the crystalline
phase is given by Eq. (\ref{rate}).
In view of the highly approximate
nature of our theory this neglect is reasonable so Eq. (\ref{rate})
is our approximation for the nucleation rate.
If $\rho_c(n_{min})$ is very small then if the constraint is removed
the distribution of cluster densities will change only very slowly.
Therefore
the unconstrained fluid will persist for a long time, much longer than
$\tau$, and so the unconstrained fluid phase is observable:
it is metastable. However, if $\rho_c(n_{min})$ is not very small then
as soon as the constraint is removed the unconstrained fluid starts
to crystallise. The unconstrained fluid does not last long enough
to be observable: it is unstable. What constitutes a very small density
is of course rather arbitrary but we will try to quantify it when we
discuss our results in the next section.

\section{Results}

Experiments on globular proteins have found metastable
\cite{debenedetti}
fluid--fluid transitions \cite{broide91,muschol97}, i.e., a
fluid-fluid transition which lies within the fluid-solid
coexistence region. The crystallisation
of proteins is often slow, taking several days, which allows
the protein solution to be cooled into a region of the phase
diagram where the fluid phase separates into two fluid phases
of differing densities.
Therefore, we show phase diagrams,
in Figs. \ref{fig2} and \ref{fig3}, in which the fluid-fluid
transition lies within the fluid-solid coexistence region.
For other values of the parameters
of the models, $n_s$, $\theta_c$ and $r_c$, there is a stable
fluid-fluid transition \cite{searprot}.
Fig. \ref{fig2} is the phase diagram of a model protein with 4 sites
and Fig. \ref{fig3} is the phase diagram for a model with 6 sites
and a larger value of $\theta_c$.
These two models were chosen as their
phase diagrams were calculated and discussed in Ref. \cite{searprot}
and they differ markedly in how deep the fluid-fluid transition is
into the fluid-solid coexistence region.
In Fig. \ref{fig2}, the fluid-fluid critical point is
at a volume fraction $\eta=0.090$ and at reciprocal
temperature $\epsilon/kT=10.24$.
We can use as a measure of how deep the fluid-fluid transition
is into the fluid-solid coexistence region the ratio of the
temperature at the critical point to that of
a fluid of the same density which coexists with the solid.
For the model of Fig. \ref{fig2} fluid at a volume fraction
$\eta=0.090$ coexists with the solid at $\epsilon/kT=8.37$.
The ratio of the temperatures is then 0.82.
For the model of Fig. \ref{fig3} the critical point is
at $\eta=0.154$ and $\epsilon/kT=7.18$. A fluid
with this density coexists with the solid phase at $\epsilon/kT=4.54$.
The ratio of the temperatures is now 0.63.
Note that our temperature is a reduced temperature, a dimensionless
ratio $kT/\epsilon$.
We have plotted our phase diagrams as a function of $kT/\epsilon$
but this scale is not directly
related to the real temperature of a protein solution
as the protein-protein interactions (which determine $\epsilon$)
vary with the temperature of the experiment.

In Figs. \ref{fig2} and \ref{fig3} we have shown as a dot-dashed
curve an estimate of where percolation occurs in the fluid.
At percolation the association of the molecules is sufficiently
strong that an infinite cluster appears \cite{degennes}, that is
to say that there are an infinite number of the molecules which are
joined to each other via pathways of bonds. The percolation curve
gives us an indication of when the density is too high or the interactions
too strong for our approximation that the crystalline clusters
interact weakly with the surrounding fluid to be valid. We will not
use our approximation for the cluster densities, Eq. (\ref{rhocon}),
beyond (i.e., to the right of) the percolation curve.
See Ref. \cite{degennes} for an introduction to percolation. If we
neglect loops of bonds we obtain what is called the classical
theory of percolation which predicts that percolation occurs at
a fraction of bonds $(1-X_p)$ given by \cite{degennes}
\begin{equation}
1-X_p=\frac{1}{n_s-1}~~~\mbox{or}~~~X_p=1-\frac{1}{n_s-1},
\label{xp}
\end{equation}
$X_p$ is the fraction of sites not bonded when percolation occurs.

Now we will use Eq. (\ref{rhocon}) to calculate
the cluster densities within the dilute fluid part of the fluid-solid
coexistence region of the phase diagrams in Figs. \ref{fig2} and
\ref{fig3}. For the phase diagram of Fig. \ref{fig2}, the 4-site model,
we have calculated cluster densities in the region of the phase
diagram bounded at the right by curve where percolation occurs, from
below by the curve describing the density of the fluid phase which
coexists with the crystal and from above by the density of the
dilute fluid phase which coexists with the dense fluid phase.
The approximations we used to calculate the cluster densities,
$\rho_c(n)$, are only reasonable at low densities and away
from a critical point. The region is bounded from above
by the fluid-fluid coexistence curve as we expect
the fluid to become
unstable with respect to condensation a little inside the
coexistence curve and so our calculated cluster densities are
meaningless there. We expect condensation to occur only a little
into the fluid-fluid coexistence region as we expect the
interfacial tension between the two fluid phases will be
small and therefore that nucleation of the dense fluid phase will be rapid
except very near the coexistence curve.

Throughout this region the densities of crystalline clusters of all
sizes $n=8$ and up are tiny. For example, at $\eta=0.1$ and
$\beta\epsilon=9$ the density of crystalline clusters
of 8 spheres is $\rho_c(8)={\cal O}(10^{-21}\sigma^{-3})$
and as $n$ increases the
density rapidly decreases. So, the density of even small
crystalline clusters is negligible.
The nucleation rate, Eq. (\ref{rate}),
is effectively zero and the dilute fluid phase will
be stable with respect to crystallisation effectively indefinitely: it
is metastable.
This finding that the crystal cannot nucleate from a dilute fluid is
interesting as experiments on solutions of many proteins find
it difficult or impossible to find crystallisation. 

The nucleation rate is so low because the nuclei,
the crystalline clusters have extremely low densities. This
can be traced to the interfacial term in our expression for
$\rho_c$, Eq. (\ref{rhocfin}). This is the second term in the
exponential which varies as the number of molecules
at the surface, as $n^{2/3}$. It is large because under conditions
that the crystal coexists with a dilute fluid the ratio between
the attraction energy and the thermal energy $\epsilon/kT$
is large.
At the surface of the cluster bonds are broken
and each broken bond decreases the density of a nucleus
by $\exp(-\beta\epsilon/2)$, which is rather large.
In the language of classical nucleation theory \cite{debenedetti}
the barrier to nucleation is high because the surface tension is high.
The surface tension $\gamma$ here comes from the energy of the broken bonds,
$\gamma\simeq(1/2)\epsilon\sigma^{-2}+\Delta a\sigma^{-2}$, where
$\Delta a$ is small, ${\cal O}(-0.1kT)$.



In view of the extremely small numbers we have not plotted cluster
densities for the model parameters of Fig. \ref{fig2}. However,
the fluid-fluid transition is deeper in the fluid-solid coexistence
region in Fig. \ref{fig3} so larger cluster densities are achievable.
Plots of $\rho_c(n)$ against $n$ for three points in the dilute phase
of Fig. \ref{fig3} are shown in Fig. \ref{fig4}. The three points
are chosen to be at roughly the highest densities at which the
theory is reliable and the fluid is outside the fluid-fluid
coexistence region the solid, dashed and dotted curves
the supersaturations $\beta(\mu-\mu_s)$ are 3.71, 4.77 and 5.52,
respectively.
An approximation to the nucleation rate is given by Eq. (\ref{rate})
which is proportional to the densities at the minima of the curves in
Fig. \ref{fig4}. We can get an estimate of what the numbers mean for
a protein solution.
Protein molecules are a few nms
in diameter so in a sample 1mm across there are of order $10^{16}$
protein molecules. At $\beta\epsilon=7.5$, $\eta=0.05$,
$\rho_c(n_{min})={\cal O}(10^{-16}\sigma^{-3})$
so in a sample 1mm across we have
${\cal O}(1)$ crystallites nucleating in the sample per time $\tau$.
Muschol and Rosenberger \cite{muschol95} estimate diffusivities
for lysozyme (a well studied globular protein) of order
$10^{-10}$m$^2$s$^{-1}$. The characteristic time of the dynamics $\tau$
should be of order the time a protein takes to diffuse its own
diameter, this time is the square of the diameter, $10^{-17}$m$^2$
divided by the diffusion constant $10^{-10}$m$^2$s$^{-1}$, so
we have $\tau={\cal O}(10^{-7}{\mbox s})$. So we end up with
the rough estimate of $10^7$ crystallites nucleating in the sample per
second. Nucleation is therefore rapid.
In common with classical nucleation theory our approximation for the
nucleation rate is the calculation of a very small number and so the
errors are typically large,
easily several orders of magnitude \cite{debenedetti}.
Bearing this in mind our theory can only tell us that the model
parameters of Fig. \ref{fig3} lie close to the dividing line
between parameter values for which nucleation of crystalline phase
from a dilute fluid phase
is not achievable on experimental time scales and parameter values
for which it is.

George and Wilson \cite{george94} determined the
second virial coefficients of a number of globular proteins under the
conditions for which they crystallised. They found that the
values of the second virial coefficients lay within a  small range, which
they called the `crystallisation slot'.
Using Eq. (\ref{b2}) for the second virial coefficient, $B_2$,
we can determine the values of $B_2$ at the 3 temperatures for
which we plotted the cluster densities in Fig. \ref{fig4}.
They are $B_2=0.21\sigma^3$, $-3.03\sigma^3$ and $-6.35\sigma^3$
for $\beta\epsilon=6$, 7 and 7.5, respectively. So, although at
all 3 temperatures we have (at different volume fractions) similar
densities of the minimum-density cluster the second virial
coefficient varies over a large range, it even changes sign.
Thus, our results for the nucleation rate do not offer an explanation of
George and Wilson's finding.

\section{Conclusion}

We have studied a simple model of a globular protein molecule
in solution.
The phase diagram and the densities of crystalline clusters in the
dilute fluid phase have been calculated. The phase diagram
predicted by our bulk free energy includes fluid-fluid
coexistence within the fluid-crystal coexistence region.
When this fluid-fluid coexistence region is not too deep into
the fluid-crystal coexistence region, as in Fig. \ref{fig2},
we find that the dilute fluid phase outside of the fluid-fluid
coexistence region is metastable, i.e., the rate of nucleation of the
crystalline phase is negligible. It is not possible to produce
a crystal directly from the dilute fluid for this model.
When the fluid-fluid coexistence region is deeper into the
fluid-crystal coexistence region, as in Fig. \ref{fig3},
the nucleation rate becomes large enough to be observable
within the dilute fluid.

Essentially, we defined the dilute fluid
as being the fluid at densities below the percolation threshold. This
means that the fluid-fluid critical point is not included in our
definition of the dilute fluid. Ten Wolde and Frenkel \cite{tenwolde97}
have shown that near a fluid-fluid critical point the interface
between the crystalline nucleus and the surrounding fluid is diffuse
and that this enhances the nucleation rate dramatically.
The diffuse interface is very different from the sharp interface
we had to assume to obtain approximations for the cluster densities
and hence the nucleation rate.
If we consider
the (highly inaccurate) predictions of our theory near the critical
points of Figs. \ref{fig2} and \ref{fig3}, we find that
$\rho_c(n_{min})={\cal O}(10^{-135}\sigma^{-3})$
and ${\cal O}(10^{-14}\sigma^{-3})$,
respectively. So, nucleation is certainly rapid near the critical
point of Fig. \ref{fig3}. However, the density $\rho_c(n_{min})$
is predicted to be so low near the critical point of Fig \ref{fig2},
that even taking into account the very large errors in our theory
we would not expect nucleation. Although the nucleation rate is
enhanced by the nature of the fluid near its critical point,
as the model parameters are varied to move the
critical point toward the fluid-crystal
coexistence curve, the rate will tend to zero.
In the limit that
the critical point touches the fluid-crystal coexistence curve, i.e.,
at the point where the fluid-fluid transition goes from being metastable
to being stable, the supersaturation at the critical point tends to
zero, reducing the nucleation rate to zero.

As this work has been motivated by the difficult and important
problem of crystallising globular proteins it is interesting
to speculate on how the model of Fig. \ref{fig2} could be
crystallised.
The nucleation rate is far too low in the dilute fluid so in
order to increase the rate the fluid must either be made more dense
or the attractions strengthened. Both of these may result in
equilibrium being difficult to reach with the result that fluid
could become gel-like.
Also, if the fluid undergoes a fluid-fluid transition
its density and hence its nucleation rate jumps \cite{muschol97}.
There is an optimum nucleation rate to obtain good, i.e., large with few
defects, crystals. Now, if there were no fluid-fluid transition then
the crystalline cluster densities and hence the nucleation rate
vary continuously but at condensation the densities will jump so there
is a risk that the nucleation rate will jump over the optimum one
making good crystals hard to obtain. 
Crystallisation would be facilitated if the free energy cost
of the surface of the cluster, the second term in the
exponential of Eq. (\ref{rhocfin}), was less. If the interactions
were less directional then the crystal would be stable at higher
temperatures, i.e., smaller values of $\beta\epsilon$, where the
surface would have a lower free energy.


\newpage
\begin{figure}
\caption{
\lineskip 6pt
\lineskiplimit 6pt
A schematic illustrating a crystalline cluster of 8 of our model
globular protein molecules. The 8 molecules are arranged at the corners
of a cube. The core of the proteins is represented by a shaded sphere
and the sites which mediate the directional attractions by black discs.
Only the sites facing us are shown. The model illustrated is the 6-site
model.
\label{fig1}
}
\begin{center}
\epsfig{file=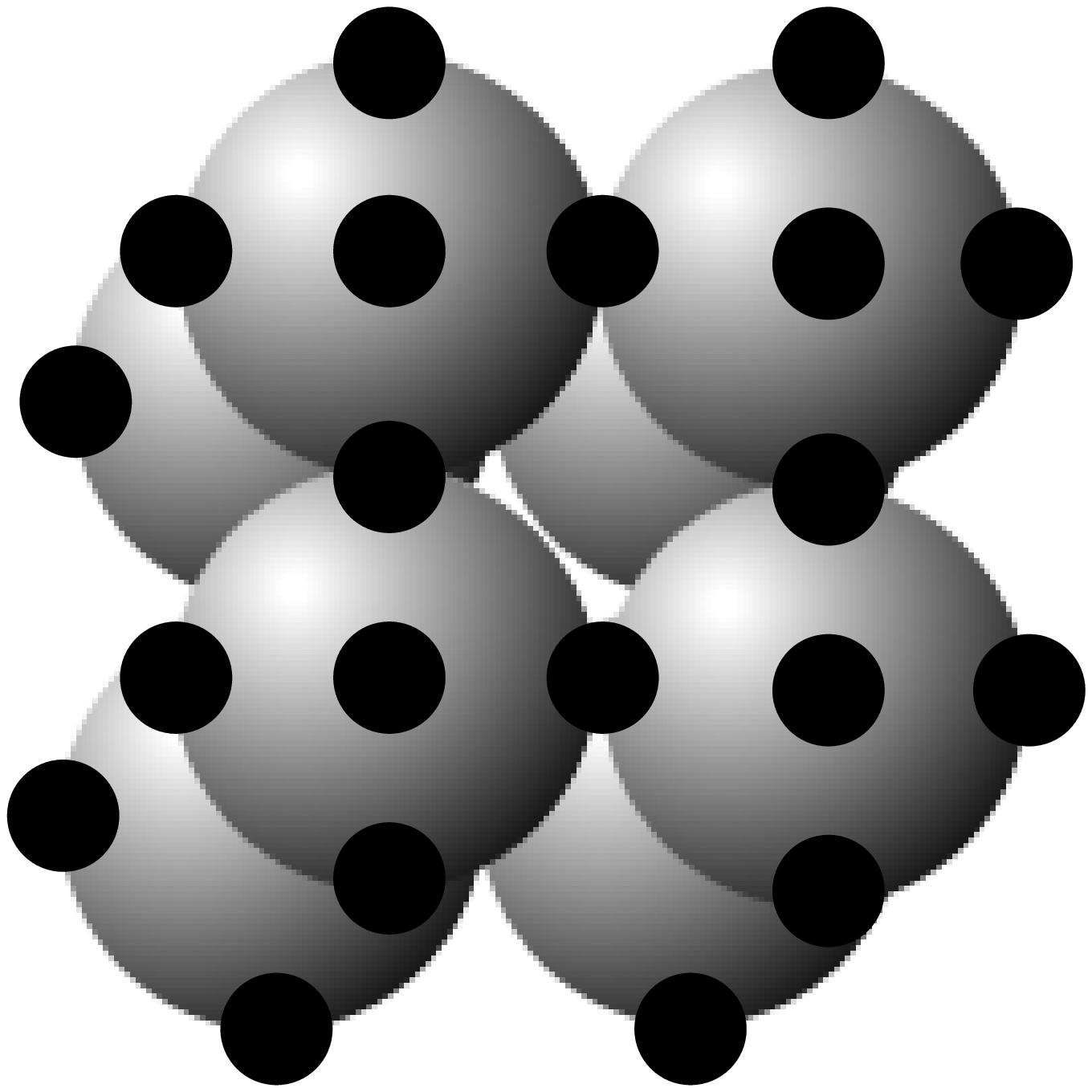,width=3.0in}
\end{center}
\end{figure}

\begin{figure}
\begin{center}
\caption{
\lineskip 6pt
\lineskiplimit 6pt
The phase diagram of our model of a globular protein. The number
of sites $n_s=4$, $r_c=1.05\sigma$ and $\theta_c=0.3$ radians
or about 17$^\circ$.
The solid curves separate the one and two-phase regions.
The letters F, S and 2 denote the regions of the phase space
occupied by the fluid phase, the solid phase and coexistence
between the fluid and solid phases. The dashed
curve is the coexistence curve for a metastable fluid--fluid
transition. The dot-dashed curve is the estimated percolation
threshold.
\label{fig2}
}
\epsfig{file=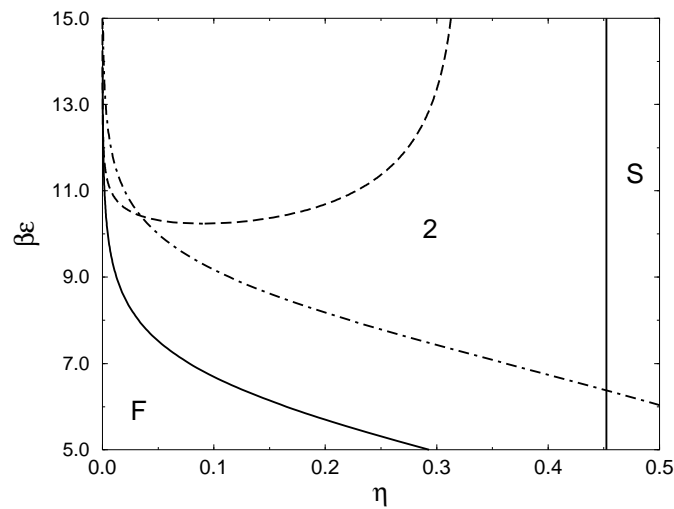,width=3.0in,angle=270}
\end{center}
\end{figure}

\begin{figure}
\begin{center}
\caption{
\lineskip 6pt
\lineskiplimit 6pt
The phase diagram of our model of a globular protein. The number
of sites $n_s=6$, $r_c=1.05\sigma$ and $\theta_c=0.45$ radians
or about 26$^\circ$. See the caption to Fig. \ref{fig2} for
the meaning of the curves.
\label{fig3}
}
\epsfig{file=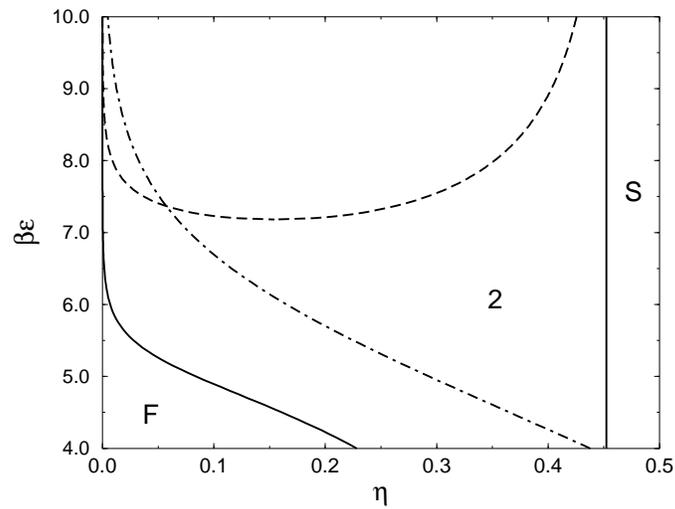,width=3.0in,angle=270}
\end{center}
\end{figure}

\begin{figure}
\begin{center}
\caption{
\lineskip 6pt
\lineskiplimit 6pt
The densities of crystalline clusters $\rho_n$,
Eq. (\ref{rhocon}), as a function of
$n$ at three points in the phase diagram of Fig. \ref{fig3}. The
solid, dashed and dotted curves are for
$\eta=0.2$, $\beta\epsilon=6$;
$\eta=0.1$, $\beta\epsilon=7$ and
$\eta=0.05$, $\beta\epsilon=7.5$, respectively.
\label{fig4}
}
\epsfig{file=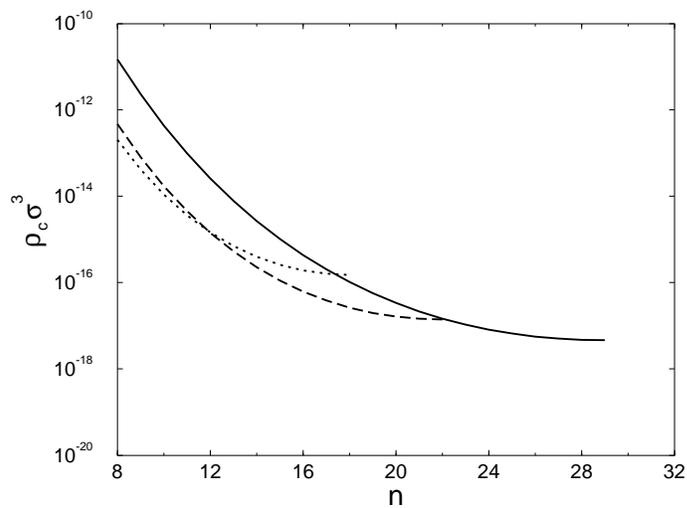,width=3.0in,angle=270}
\end{center}
\end{figure}

\end{document}